\documentclass[tightenlines,twoside,secnumarabic,superscriptaddress,preprintnumbers,
               onecolumn,floatfix,nofootinbib,showpacs,11pt]{revtex4}

\usepackage[english]{babel}
\usepackage{amsmath,amssymb}
\usepackage{graphicx}
\usepackage{natbib}
\usepackage[hypertex]{hyperref}

\begin{document}

%==============================================================================
\title{Interpretation of MINOS data in terms of non-standard neutrino interactions}
\author{Joachim Kopp}        \email[Email: ]{jkopp@fnal.gov}
\affiliation{Theoretical Physics Department, Fermi National Accelerator Laboratory,
             P.O.~Box 500, Batavia, IL 60510, USA}
\author{Pedro A.~N.~Machado} \email[Email: ]{accioly@fma.if.usp.br}
\affiliation{Instituto de F\'{i}sica, Universidade de S\~{a}o Paulo,
             C.P.~66.318, 05315-970 S\~{a}o Paulo, Brazil}
\affiliation{Theoretical Physics Department, Fermi National Accelerator Laboratory,
             P.O.~Box 500, Batavia, IL 60510, USA}
\author{Stephen J.~Parke}    \email[Email: ]{parke@fnal.gov}
\affiliation{Theoretical Physics Department, Fermi National Accelerator Laboratory,
             P.O.~Box 500, Batavia, IL 60510, USA}
\pacs{14.60.Pq, 12.60.-i, 13.15.+g}
\preprint{FERMILAB-PUB-10-238-T}
\preprint{INT-PUB-10-038}
\date{Sep 1, 2010}
%==============================================================================

\begin{abstract}
  The MINOS experiment at Fermilab has recently reported a tension between the
  oscillation results for neutrinos and anti-neutrinos. We show that this
  tension, if it persists, can be understood in the framework of non-standard
  neutrino interactions (NSI). While neutral current NSI (non-standard matter
  effects) are disfavored by atmospheric neutrinos, a new charged
  current coupling between tau neutrinos and nucleons can fit the MINOS data
  without violating other constraints. In particular, we show that loop-level
  contributions to flavor-violating $\tau$ decays are sufficiently
  suppressed.  However, conflicts with existing bounds could arise once the
  effective theory considered here is embedded into a complete renormalizable
  model.  We predict the future sensitivity of the T2K and NO$\nu$A experiments
  to the NSI parameter region favored by the MINOS fit, and show that both
  experiments are excellent tools to test the NSI interpretation of the MINOS
  data.
\end{abstract}

%==============================================================================
\maketitle
%==============================================================================

%==============================================================================
\section{Introduction}
%==============================================================================

Recently the Fermilab MINOS experiment has reported new results on
$\bar{\nu}_\mu$ disappearance~\cite{MINOS:Nu2010}.  Interestingly, the values
of the neutrino oscillation parameters $\Delta m^2_{32}$ and $\sin^2
2\theta_{23}$ preferred by this anti-neutrino measurement are in tension, at
the 90\% confidence level, with the preferred region in the $\Delta m^2_{32}$
versus $\sin^2 2\theta_{23}$ plane for the neutrino $\nu_\mu$ disappearance.  A
likely explanation of this tension is lack of statistics especially in the
anti-neutrino channel where the number of observed events in the far detector
is approximately 100 whereas for the neutrino channel the number of events in
the far detector is of order 20 times larger.  One could speculate that the
tension between neutrino and anti-neutrino disappearance might be the first
hint  of CPT violation in the neutrino sector~\cite{Dighe:2008bu, Diaz:2009qk,
Barenboim:2009ts, Choudhury:2010vj}. However, given that CPT conservation is
such an important tenet of modern quantum field theory, it is important to
explore other possibilities for new physics before giving up CPT conservation.
Since the MINOS experiment is performed not in vacuum but with more than 700
kilometers of Earth matter between the source and the detector, there is the
possibility of Wolfenstein type matter effects~\cite{Wolfenstein:1977ue}
leading to \emph{apparent} CPT violation. While matter effects are not relevant
to $\nu_\mu$ disappearance in MINOS in the standard three-flavor framework,
they may become important in scenarios with sterile
neutrinos~\cite{Engelhardt:2010dx}, or if new non-standard interactions are
contributing to the potential that the neutrinos experience when traveling
through matter. The latter option will be explored in this paper. In addition,
we will also consider non-standard interactions modifying the neutrino
detection process in a CP non-conserving way.

Non-Standard Interactions (NSI) in the neutrino sector have been introduced
first as an \emph{alternative} to standard
oscillations~\cite{Wolfenstein:1977ue}, and later as a possible
addition~\cite{Grossman:1995wx}. The phenomenology of such subdominant NSI
effects in current and near-future accelerator neutrino experiments has been
investigated by many authors~\cite{Ota:2001pw, Ota:2002na, Kitazawa:2006iq,
Friedland:2006pi, Blennow:2007pu, Kopp:2007ne, Ribeiro:2007jq, Blennow:2008ym},
and bounds on NSI have been derived from oscillation and non-oscillation
data~\cite{Fornengo:2001pm, Davidson:2003ha, GonzalezGarcia:2007ib,
Biggio:2009nt}. In the context of the latest MINOS results, NSI have been
brought up in~\cite{Mann:2010jz, Akhmedov:2010vy}, and a concrete model has
been proposed in~\cite{Heeck:2010pg}.

In the following, we will first show analytically how NSI affect neutrino
oscillations in the two flavor limit (sec.~\ref{sec:analytic}), and then
perform fits to the MINOS $\nu_\mu$ and $\bar{\nu}_\mu$ data including
different types of NSI (sec.~\ref{sec:fit}). In sec.~\ref{sec:future}, we will
consider the potential of the T2K~\cite{T2K:2006} and
NO$\nu$A~\cite{Ayres:2004js} experiments to test the NSI interpretation of
MINOS.  Finally, we will discuss our results from the model-building point of
view and draw our conclusions in sec.~\ref{sec:conclusions}.

%==============================================================================
\section{Analytical framework of non-standard interactions}
\label{sec:analytic}
%==============================================================================

%------------------------------------------------------------------------------
\subsection{Neutral current NSI}
\label{sec:analytic-NC}
%------------------------------------------------------------------------------

At $\mathcal{O}(\text{GeV})$ energies relevant to neutrino oscillation experiments,
non-standard interactions can be introduced in the Lagrangian as effective
dimension 6 operators coupling neutrinos and charged fermions.
We will first discuss new neutral current couplings of $\mu$ and $\tau$
neutrinos to normal matter, i.e.\ electrons, up-quarks, and down-quarks.
Phenomenologically, such operators will result in non-standard matter effects,
so that the Hamiltonian governing neutrino propagation in the $\mu$--$\tau$
sector will read
\begin{align}
  H = \frac{1}{2 E} \bigg[
      U \begin{pmatrix}
           0 & \\
             & \Delta m_{32}^2
         \end{pmatrix} U^\dag + 
      A \begin{pmatrix}
          \epsilon^m_{\mu\mu}     & \epsilon^m_{\mu\tau} \\
          \epsilon^{m*}_{\mu\tau} & \epsilon^m_{\tau\tau}
        \end{pmatrix} \bigg] \,,
\end{align}
where
\begin{align}
  U = \begin{pmatrix}
        \cos\theta_{23} & \sin\theta_{23} \\
       -\sin\theta_{23} & \cos\theta_{23}
      \end{pmatrix}
\end{align}
is the leptonic mixing matrix with the mixing angle $\theta_{23}$, $E$ is the
neutrino energy, and $A = 2\sqrt{2} G_F N_e E$ is the matter potential depending
on the electron number density $N_e$ along the neutrino trajectory.  The
parameters $\epsilon^m_{\mu\mu}$, $\epsilon^m_{\mu\tau}$, and $\epsilon^m_{\tau\tau}$
give the relative strength of the non-standard interactions compared to
Standard Model weak interactions. The superscript $m$ indicates that these parameters
describe non-standard neutrino matter effects. $\epsilon^m_{\mu\tau}$ can in general be
complex, while $\epsilon^m_{\mu\mu}$ and $\epsilon^m_{\tau\tau}$ have to be real in
order to preserve the hermiticity of the Hamiltonian.  In the following, we will
set $\epsilon^m_{\mu\mu} = 0$ since terms proportional to the identity matrix do
not affect the neutrino oscillation probability, implying that oscillation
experiments are only sensitive to the combination $\epsilon^m_{\tau\tau} -
\epsilon^m_{\mu\mu}$.

The disappearance probability for $\nu_\mu$ in matter of constant density is
given by
\begin{eqnarray}
  P(\nu_\mu \rightarrow \nu_\mu) =
    1 - \sin^2 2\theta_N \sin^2 \left(\frac{\Delta m^2_N L}{4E}\right) \,.
\end{eqnarray}
where $\theta_N$ and $\Delta m^2_N$ are the mixing angle and mass squared
difference in matter.  These matter oscillation parameters can be found by
solving the following set of coupled equations,
\begin{eqnarray}
  \Delta m^2_N \cos 2\theta_N
    &=&  \Delta m^2_{32} \cos 2 \theta_{23} + \epsilon^m_{\tau\tau}A \,, \\
  \Delta m^2_N \sin 2\theta_N e^{i\phi_N}
    &=&  \Delta m^2_{32} \sin 2 \theta_{23} + 2\epsilon^m_{\mu\tau}A \,.
\end{eqnarray}
The phase $\phi_N$ is unobservable in oscillations and can be phased away.
The solution is trivial to find and is given by
\begin{eqnarray}
  \Delta m^2_N &=&
    \sqrt{ (\Delta m^2_{32} \cos 2 \theta_{23} + \epsilon^m_{\tau\tau}A)^2
          + \vert \Delta m^2_{32} \sin 2 \theta_{23} + 2 \epsilon^m_{\mu\tau}A\vert ^2 } \,
  \label{eqn:dmsq} \\[0.25cm]
  {\rm and} \quad \sin^2 2\theta_N &=&
    \vert\Delta m^2_{32} \sin 2 \theta_{23} + 2\epsilon^m_{\mu\tau}A\vert^2/\Delta m^4_N \,.
  \label{eqn:ssqth}
\end{eqnarray}
Thus, the disappearance probability can be written as
\begin{eqnarray}
  P(\nu_\mu \rightarrow \nu_\mu) &=&
    1 - \frac{\vert\Delta m^2_{32} \sin 2 \theta_{23} + 2\epsilon^m_{\mu\tau}A\vert^2}{\Delta m^4_N}
        \sin^2 \left(\frac{\Delta m^2_N L}{4E}\right) \,.
  \label{eq:P-NC}
\end{eqnarray}
with $\Delta m^2_N$ given by eq.~\eqref{eqn:dmsq}.   This expression agrees
with that found in ref.~\cite{Kopp:2007ne} when expanded to first order in
$\epsilon^m_{\mu\tau}$ and $\epsilon^m_{\tau\tau}$.

In vacuum, $A=0$, and eq.~\eqref{eq:P-NC} reduces to the standard two flavor
disappearance probability.  In the small $L/E$  limit, i.e.\ when $\sin^2
(\Delta m^2_N L / 4 E) \approx (\Delta m^2_N L / 4E)^2$,
\begin{eqnarray}
  P(\nu_\mu \rightarrow \nu_\mu) &\approx&
    1- \left| \sin 2 \theta_{23} + \frac{2\epsilon^m_{\mu\tau}A}{\Delta m^2_{32}}\right|^2
       \left(\frac{\Delta m^2_{32}L}{4E}\right)^2 \,,
\end{eqnarray}
so  $\epsilon^m_{\mu\tau}$ modifies the disappearance probability in this limit
whereas $\epsilon^m_{\tau\tau}$ does not.

For anti-neutrinos, $\epsilon^m_{\mu\tau} \rightarrow \epsilon^{m*}_{\mu\tau}$ and
$A \rightarrow -A$, so that in matter
\begin{eqnarray}
P(\nu_\mu \rightarrow \nu_\mu) \neq P(\bar{\nu}_\mu \rightarrow \bar{\nu}_\mu) 
\end{eqnarray}
without CPT violation.  

We note three interesting symmetries in eq.~\eqref{eq:P-NC}: First, the
expression depends only on $\cos[\arg(\epsilon^m_{\mu\tau})]$, not on
$\sin[\arg(\epsilon^m_{\mu\tau})]$.  Therefore, it is invariant under the
replacement
\begin{align}
  \arg(\epsilon^m_{\mu\tau}) \to 2\pi n - \arg(\epsilon^m_{\mu\tau})
  \label{eq:symm1-NC}
\end{align}
for arbitrary integer $n$.  Moreover, it is easy to verify that $P(\nu_\mu
\rightarrow \nu_\mu)$ is also invariant under the simultaneous replacements
\begin{align}
  \epsilon^m_{\mu\tau}  \to -\epsilon^m_{\mu\tau}  \,, \qquad
  \epsilon^m_{\tau\tau} \to -\epsilon^m_{\tau\tau} \,, \qquad
  \Delta m_{32}^2       \to -\Delta m_{32}^2 \,.
  \label{eq:symm2-NC}
\end{align}
as well as under the transformation
\begin{align}
  \epsilon^m_{\tau\tau} \to -\epsilon^m_{\tau\tau} \,, \qquad
  \theta_{23}           \to \frac{\pi}{2} - \theta_{23} \,.
  \label{eq:symm3-NC}
\end{align}
These symmetries will generate an eightfold degeneracy.  For fixed $E$,
$P(\nu_\mu \rightarrow \nu_\mu)$ has an additional, continuous, symmetry: It is
invariant under any simultaneous variation of $\Delta m_{32}^2$, $\theta_{23}$
, $\epsilon^m_{\tau\tau}$, $|\epsilon^m_{\mu\tau}|$, and
$\arg(\epsilon^m_{\mu\tau})$ that leaves $\Delta m_N^2$ and $\sin^2 2\theta_N$
invariant. If we demand this invariance for neutrinos and anti-neutrinos, we
obtain 4 equations for 5 free parameters, implying that the symmetry is
continuous.  However, since $A$ is energy-dependent, this symmetry will not be
manifest in the MINOS data, which covers a broad range of energies, and we will
therefore not consider it further in this paper.

Note that these symmetries are exact only in the two-flavor framework. In the
three-flavor case with large $\theta_{13}$, it is for example possible to
determine the mass hierarchy by observing \emph{standard} matter effects either
in the $\nu_\mu \to \nu_e$ channel or directly in the $\nu_\mu \to \nu_\mu$
channel~\cite{Gandhi:2004md}.

\begin{figure}
  \vspace*{-2cm}
  \begin{center}
    \includegraphics[width=15cm]{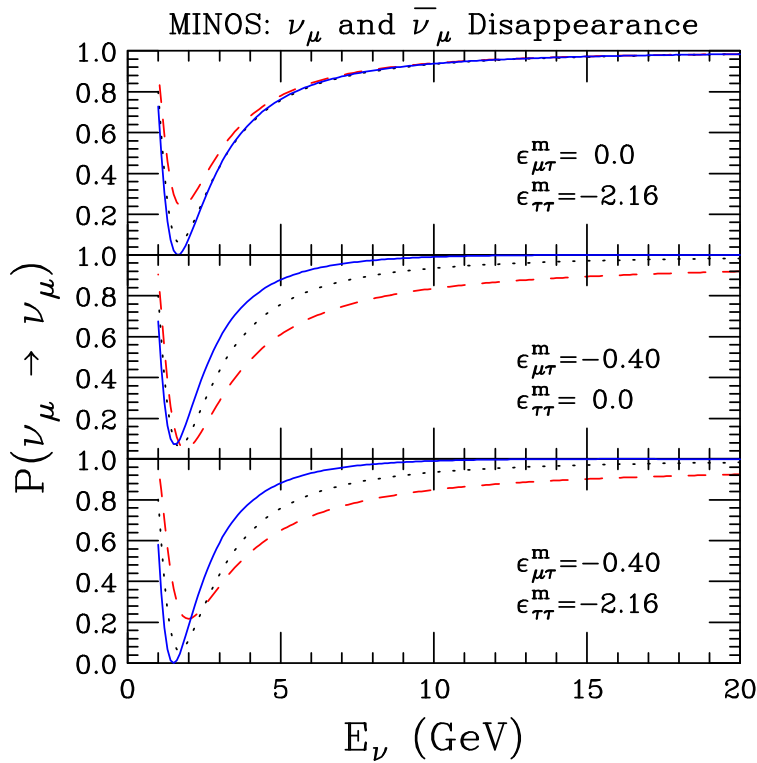}
  \end{center}
  \vspace*{-7cm}
  \caption{The survival probability for MINOS ($L = 735$~km) with a variety of
    neutral current NSI (non-standard matter effects) turned on as indicated in
    the plot. The solid (blue) lines are the neutrino survival probabilities
    whereas the dashed (red) lines are for anti-neutrinos.  The dotted (black)
    lines are the vacuum survival probabilities.  For the standard oscillation
    parameters, we have assumed $\Delta m_{32}^2 = +2.86 \times 10^{-3}$~eV$^2$
    and $\sin^2 \theta_{23} = 0.38$.}
  \label{fig:P-NC}
\end{figure}

In fig.~\ref{fig:P-NC} we have plotted the disappearance probabilities with
all three combinations of non-zero  $\epsilon^m_{\mu\tau}$ and/or
$\epsilon^m_{\tau\tau}$ for representative values of the $\epsilon^m$'s for the
MINOS experiment.  One can see that non-zero $\epsilon^m_{\mu\tau}$ changes the
disappearance probability most notably at large energies and shifts the
position of the minimum in energy. Whereas non-zero $\epsilon^m_{\tau\tau}$
changes the disappearance probability most notably near the first oscillation
minimum, especially in the depth of the minimum.  Since the tension between
MINOS neutrino and anti-neutrino data is both in the position of the minimum
and in its depth, one requires non-zero $\epsilon^m_{\mu\tau}$ and
non-zero $\epsilon^m_{\tau\tau}$ in order to lift the tension in the
optimal way.

%------------------------------------------------------------------------------
\subsection{Charged current NSI}
\label{sec:analytic-CC}
%------------------------------------------------------------------------------

As an alternative to neutral current NSI, we also discuss non-standard charged
current interactions affecting the neutrino production and/or detection
processes as an explanation for the MINOS results. If the Wilson coefficients
of the corresponding effective operators are complex, the interference term
between the standard and non-standard Feynman amplitudes can be different for
neutrinos and anti-neutrinos and CP-violating phenomena can emerge.  The
modifications to the far detector event spectra observed in MINOS can be
induced by (i) operators leading to a modified flux of $\nu_\mu$ at the far
detector, but not at the near detector, and (ii) by operators leading to the
production of muons in interactions of $\nu_\tau$. (We neglect the possibility
of non-standard interactions of $\nu_e$ since their flux at the far detector is
between one and two order of magnitude smaller than that of $\nu_\mu$ because
of the low $\nu_e$ contamination of the NuMI beam and the smallness of the
mixing angle $\theta_{13}$.) The only way of realizing case (i) in a
three-flavor framework is to postulate a $\nu_\tau$ contamination in the NuMI
beam, which would be invisible to the near detector, but would have partly
oscillated into $\nu_\mu$ when reaching the far detector. However, results from
the NOMAD experiment~\cite{Astier:2001yj} constrain the $\nu_\tau$
contamination of a NuMI-like neutrino beam at short baseline to be less than
$1.7 \times 10^{-4}$ at 90\% confidence level, much too small to be relevant to
MINOS. We therefore neglect case (i) in the following, and focus on case (ii),
a non-standard interaction of the type
\begin{align}
  \nu_\tau + N \to X + \mu \,,
\end{align}
where $N$ is a nucleon and $X$ stands for the hadronic interaction products.
The operator generating this process has the structure
\begin{align}
  \mathcal{L}_{\rm NSI} \supset -2\sqrt{2} G_F \epsilon^d_{\tau\mu} V_{ud} \,
  [\bar{u} \gamma^\rho d] \, [\bar{\mu} \gamma_\rho P_L \nu_\tau] + h.c.\,,
  \label{eq:L-NSI}
\end{align}
where $u$ and $d$ denote the up- and down-quark fields, $G_F$ is the Fermi
constant, $P_L = (1 - \gamma^5)/2,$ and $\epsilon^d_{\tau\mu}$ gives the
strength of the non-standard interaction compared to Standard Model weak
interactions. (In the notation from ref.~\cite{Biggio:2009nt}, this coefficient
would be called $\epsilon^{ud\,V}_{\mu\tau}$.) In principle, one could also
consider the axial-vector operator $[\bar{u} \gamma^\rho \gamma^5 d] \,
[\bar{\mu} \gamma_\rho P_L \nu_\tau]$, but, being parity-odd, this operator
would lead to the decay $\pi \to \mu \nu_\tau$, which is strongly constrained
by NOMAD, as discussed above. We will therefore neglect axial-vector NSI.  We
will also not consider scalar, pseudo-scalar, and tensor operators, since they
could lead to the required interference between standard- and non-standard
Feynman amplitudes only if the outgoing muon in the detector undergoes a
helicity flip~\cite{Kopp:2007ne}. For $\mathcal{O}(\text{GeV})$ muons, this
would correspond to a suppression of the non-standard interaction rate by
$m_\mu / E \sim 0.1$.

If $\epsilon^d_{\tau\mu}$ is non-zero, the counting rate in MINOS is no longer
proportional to the standard survival probability $P(\nu_\mu \to \nu_\mu)$, but
rather to an \emph{apparent} $\nu_\mu$ survival probability $\tilde{P}(\nu_\mu
\to \nu_\mu)$, defined by the number of muons produced in the detector in
interactions of neutrinos with a given energy $E$, divided by the number of
muons that would be produced in the absence of neutrino oscillations and
non-standard interactions. Thus, $\tilde{P}(\nu_\mu \to \nu_\mu)$ includes the
possibility that the neutrino flavor has changed into $\nu_\tau$ during
propagation, but non-standard interactions still lead to a final state muon,
normally associated with $\nu_\mu$ interactions. Since the amplitudes for the
processes $\nu_\mu + N \to X + \mu$ and $\nu_\mu \xrightarrow{\text{osc.}}
\nu_\tau + N \to X + \mu$ can interfere, $\tilde{P}(\nu_\mu \to \nu_\mu)$ can
be larger than unity and is therefore not a survival probability in the
usual sense.

In the two-flavor approximation we find for
$\tilde{P}(\nu_\mu \to \nu_\mu)$~\cite{Kopp:2007ne}:
\begin{align}
  \tilde{P}(\nu_\mu \to \nu_\mu) = 1
  & - \bigg[1 + 2 \, |\epsilon^d_{\tau\mu}| \cot 2\theta_{23}
                \cos\big[\arg(\epsilon^d_{\tau\mu})\big]
        - |\epsilon^d_{\tau\mu}|^2 \bigg] \sin^2 2\theta_{23}
          \sin^2\left(\frac{\Delta m^2_{32}L}{4E}\right) \nonumber\\
  & + 2 \, |\epsilon^d_{\tau\mu}| \sin 2\theta_{23}
          \sin\big[\arg(\epsilon^d_{\tau\mu})\big]
          \sin\left(\frac{\Delta m^2_{32}L}{4E}\right)
          \cos\left(\frac{\Delta m^2_{32}L}{4E}\right) \,.
  \label{eq:P-CC}
\end{align}
For anti-neutrinos, the sign of $\arg(\epsilon^d_{\tau\mu})$ has to be
reversed, and thus
\begin{eqnarray}
  \tilde{P}(\nu_\mu \rightarrow \nu_\mu) \neq
    \tilde{P}(\bar{\nu}_\mu \rightarrow \bar{\nu}_\mu) \,.
\end{eqnarray}
without CPT violation.

\begin{figure}
  \vspace*{-2cm}
  \begin{center}
    \includegraphics[width=12cm]{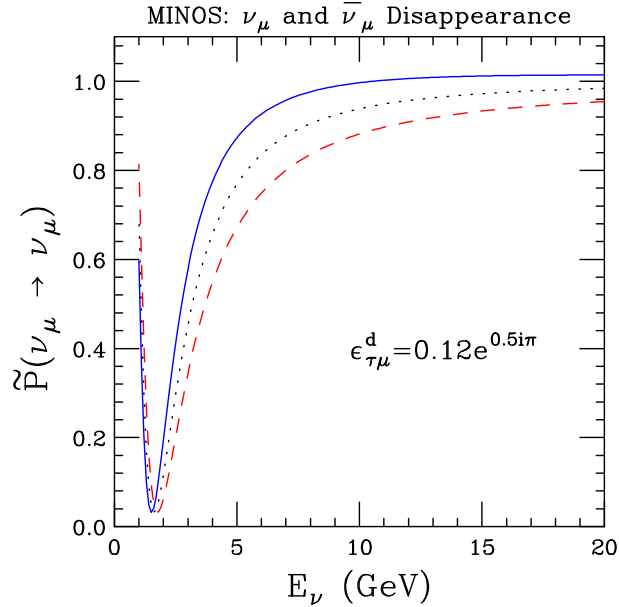}
  \end{center}
  \vspace*{-5.5cm}
  \caption{The apparent survival probability for MINOS ($L = 735$~km) without
    NSI (black dotted curve), and in a scenario with non-zero
    $\epsilon^d_{\tau\mu}$. The solid (blue) line in this case is for neutrinos,
    while the red (dashed) line is for anti-neutrinos.  For the standard
    oscillation parameters, we have assumed $\Delta m_{32}^2 = +2.74 \times
    10^{-3}$~eV$^2$ and $\sin^2 \theta_{23} = 0.41$.}
  \label{fig:P-CC}
\end{figure}

Note that the non-standard terms proportional to $|\epsilon^d_{\tau\mu}|$
and $|\epsilon^d_{\tau\mu}|^2$ in the first line of eq.~\eqref{eq:P-CC} have
the same energy dependence as the standard oscillation term and can therefore
change only the depth of the oscillation dip, but not its position. They can thus
only change the apparent value of $\sin^2 2\theta_{23}$ that would be
reconstructed in a standard oscillation analysis neglecting NSI. Moreover, these
terms do not depend on the sign of $\arg(\epsilon^d_{\tau\mu})$ and can
therefore not introduce an asymmetry between neutrinos and anti-neutrinos.  The
interference term between standard and non-standard amplitudes
in the second line of eq.~\eqref{eq:P-CC}, on the other hand, can be
different for neutrinos and anti-neutrinos. Since eq.~\eqref{eq:P-CC} is invariant under
the simultaneous replacements $\arg(\epsilon^d_{\tau\mu}) \to
-\arg(\epsilon^d_{\tau\mu})$ and $\Delta m^2_{32}L / 4E \to \pi - \Delta
m^2_{32}L / 4E$, the depth of the oscillation minimum will be the same for
neutrinos and anti-neutrinos, and only its position will be different. This is
also illustrated in fig.~\ref{fig:P-CC}, where we plot $\tilde{P}(\nu_\mu
\to \nu_\mu)$ including non-zero $\epsilon^d_{\tau\mu}$ for neutrinos and
anti-neutrinos.

Like eq.~\eqref{eq:P-NC} for neutral current NSI, also eq.~\eqref{eq:P-CC} exhibits
several symmetries. In particular, the expression is invariant under the simultaneous
replacements
\begin{align}
  \arg(\epsilon^d_{\tau\mu}) \to 2\pi n - \arg(\epsilon^d_{\tau\mu}) \,,\qquad
  \Delta m_{32}^2            \to -\Delta m_{32}^2
  \label{eq:symm1-CC}
\end{align}
and under the transformation
\begin{align}
  \arg(\epsilon^d_{\tau\mu}) \to (2n+1) \pi - \arg(\epsilon^d_{\tau\mu}) \,,\qquad
  \theta_{23}           \to \frac{\pi}{2} - \theta_{23} \,
  \label{eq:symm2-CC}
\end{align}
for arbitrary integer $n$.  Actually, the second of these symmetries can be
generalized to a continuous symmetry. To see this, note that
eq.~\eqref{eq:P-CC} is invariant under simultaneous changes of
$|\epsilon^d_{\tau\mu}|$, $\arg(\epsilon^d_{\tau\mu})$, and $\theta_{23}$,
provided that the coefficients of the energy dependent factors $\sin^2 [ \Delta
m_{32}^2 L / 4E ]$ and $\sin [ \Delta m_{32}^2 L / 4E ] \cos [ \Delta m_{32}^2
L / 4E ]$ remain invariant. This requirement imposes two constraints on the
three parameters $|\epsilon^d_{\tau\mu}|$, $\arg(\epsilon^d_{\tau\mu})$,
and $\theta_{23}$, so that there will be an infinite set of solutions.  Note
that the symmetries of eq.~\eqref{eq:P-CC}, like those of eq.~\eqref{eq:P-NC},
are exact only in the two-flavor framework.

%==============================================================================
\section{Fit to MINOS data}
\label{sec:fit}
%==============================================================================

To test the compatibility of MINOS data with the hypothesis of non-standard
neutrino interactions and to extract the allowed values for the NSI parameters,
we have performed fits using a modified version of GLoBES~\cite{Huber:2004ka,
Huber:2007ji} (see appendix~\ref{sec:sim} for details). We have first checked
that we were able to reproduce the results of the two-flavor standard
oscillation fits performed by the MINOS collaboration. We find that the fit to
only $\nu_\mu$ ($\bar{\nu}_\mu$) data yields $\chi^2/\text{dof} = 12.3/12$
($2.3/3$), while a combined fit to both data sets results in $\chi^2/\text{dof}
= 20.1/17$. According to the parameter goodness-of-fit test described in
ref.~\cite{Maltoni:2003cu}, this means that the probability for the apparent
inconsistency between the $\nu_\mu$ and $\bar{\nu}_\mu$ data sets to be merely
a statistical fluctuation is about 6\%. Even though this probability is still
relatively large, let us now study how the fit improves if the possibility of
non-standard interactions are included.

%------------------------------------------------------------------------------
\subsection{Neutral current NSI}
\label{sec:fit-NC}
%------------------------------------------------------------------------------

\begin{figure}
  \begin{center}
    \includegraphics[width=\textwidth]{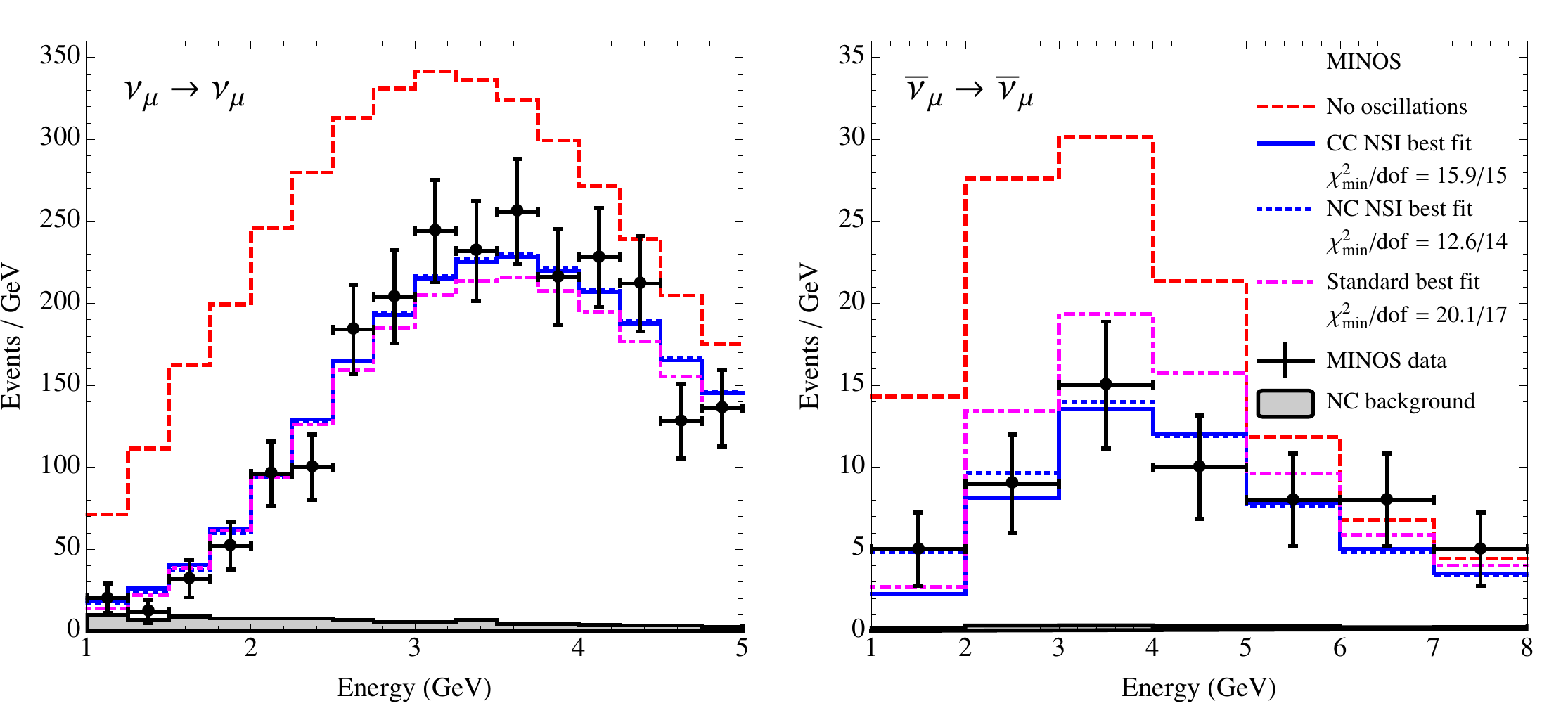}
  \end{center}
  \vspace{-.5cm}
  \caption{Comparison of MINOS data (black dots and error bars) to theoretical
    predictions including neutral-current NSI parameterized by the best fit point
    eq.~\eqref{eq:bf-NC} (blue dotted histograms) and charged current NSI
    parameterized by the best fit point eq.~\eqref{eq:bf-CC} (blue solid
    histograms).  For comparison, the red dashed histograms show the theoretical
    prediction in the absence of neutrino oscillations, and the pink dash-dotted
    histograms represent the results of a two-flavor standard oscillation fit to
    the combined $\nu_\mu$ and $\bar{\nu}_\mu$ data.}
  \label{fig:spectrum}
\end{figure}

We begin by assuming only neutral current NSI (see sec.~\ref{sec:analytic-NC}).
Using a three-flavor fit including the NSI parameters $|\epsilon^m_{\mu\tau}|$,
$\arg(\epsilon^m_{\mu\tau})$, and $\epsilon^m_{\tau\tau}$, we find the
following parameter values for the best fit point:
\begin{align}
  \begin{aligned}
    \epsilon^m_{\mu\tau}  &= -0.40 = 0.40 \, e^{1.0 i \pi} &\qquad
    \sin^2 \theta_{23}    &= \phantom{-}0.38                   \\
    \epsilon^m_{\tau\tau} &= -2.16                             &\qquad
    \Delta m_{32}^2       &= +2.86 \times 10^{-3}\ \text{eV}^2 
  \end{aligned}
  \label{eq:bf-NC}
\end{align}
The $\chi^2$ at the best fit point is 12.6, and the number of degrees of
freedom is 14. Note that the best fit value for $\epsilon^m_{\mu\tau}$ is
essentially real and negative.  An equally good fit is obtained when the sign of
$\arg(\epsilon^m_{\mu\tau})$ is inverted (eq.~\eqref{eq:symm1-NC}), when the
signs of $\epsilon^m_{\mu\tau}$, $\epsilon^m_{\tau\tau}$, and $\Delta m_{32}^2$
are flipped simultaneously (eq.~\eqref{eq:symm2-NC}), or when the sign of
$\epsilon^m_{\tau\tau}$ and the octant of $\theta_{23}$ are changed
(eq.~\eqref{eq:symm3-NC}).  This means that three-flavor effects are not large
enough to spoil the symmetries of the two-flavor survival probability
eq.~\eqref{eq:P-NC}.  Thus, the best-fit point is eightfold degenerate.  In
fig.~\ref{fig:spectrum}, we compare the theoretically predicted event spectrum
at the parameter point eq.~\eqref{eq:bf-NC} (blue dotted histograms) to the
MINOS data and to the spectra obtained from a two-flavor standard oscillation
fit.

Our fit is in qualitative agreement with the one in ref.~\cite{Mann:2010jz},
but points to somewhat larger values of $|\epsilon^m_{\mu\tau}| \sim 0.4$
compared to $|\epsilon^m_{\mu\tau}| \sim 0.1$ in ref.~\cite{Mann:2010jz}. We
surmise that the main reason for this disagreement is that our fit is based on
a simulation of the MINOS event spectrum, while, to our understanding, the
authors of ref.~\cite{Mann:2010jz} have directly fitted the oscillation
probability to the ratio of observed to expected event numbers, as computed by
the MINOS collaboration. A fit at the level of the oscillation probability,
however, cannot fully include experimental energy resolution effects.
The fit to the spectrum used in our work is able to reproduce
the MINOS best fit points as well as the allowed regions in the $\Delta
m_{32}^2$--$\sin^2 2\theta_{23}$ plane rather well (see fig.~\ref{fig:std-osc-fit}
in appendix~\ref{sec:sim}).

\begin{figure}
  \begin{center}
    \begin{tabular}{cc}
      \includegraphics[width=0.45\textwidth]{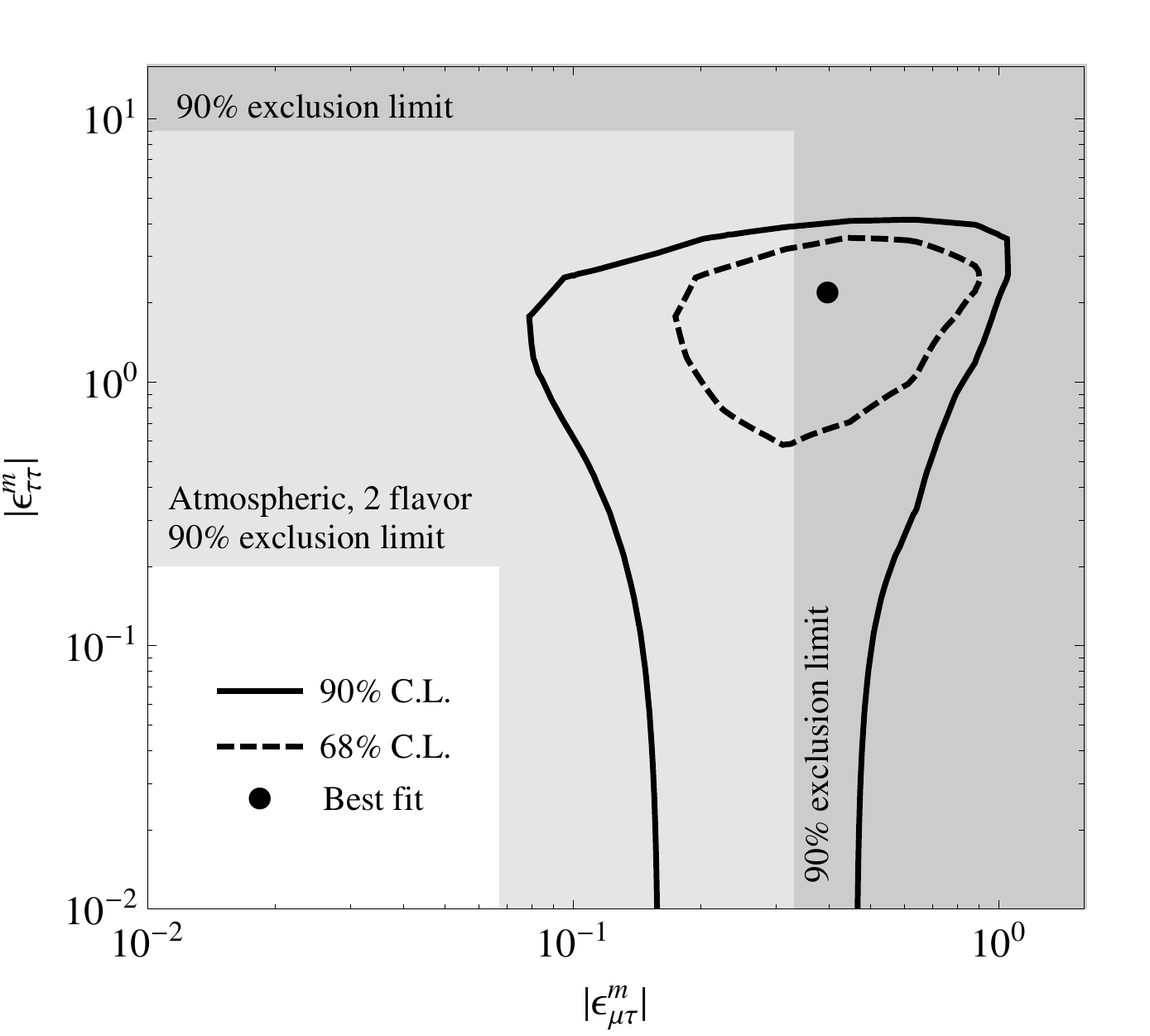} & \\
      \includegraphics[width=0.45\textwidth]{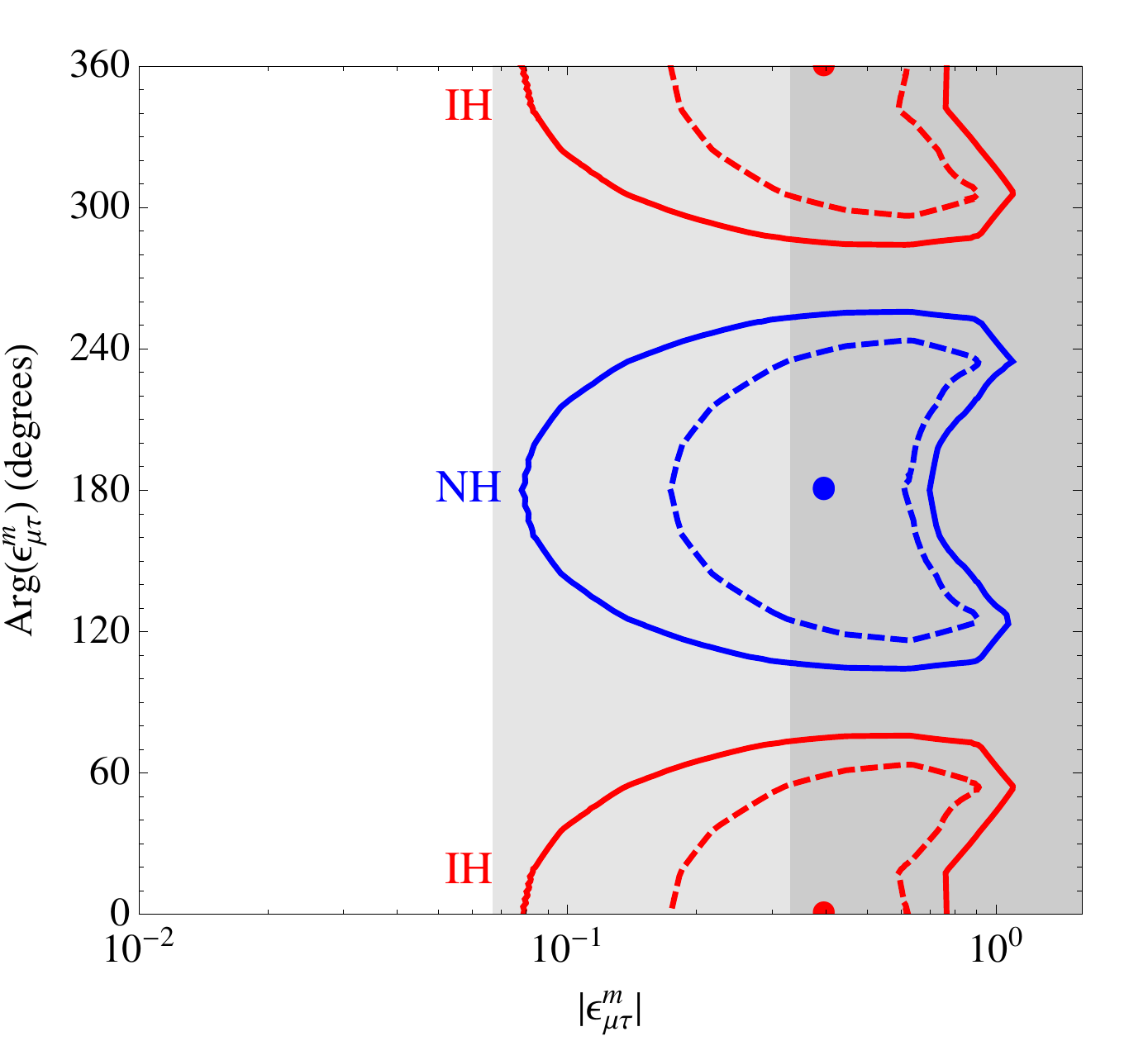} &
      \includegraphics[width=0.45\textwidth]{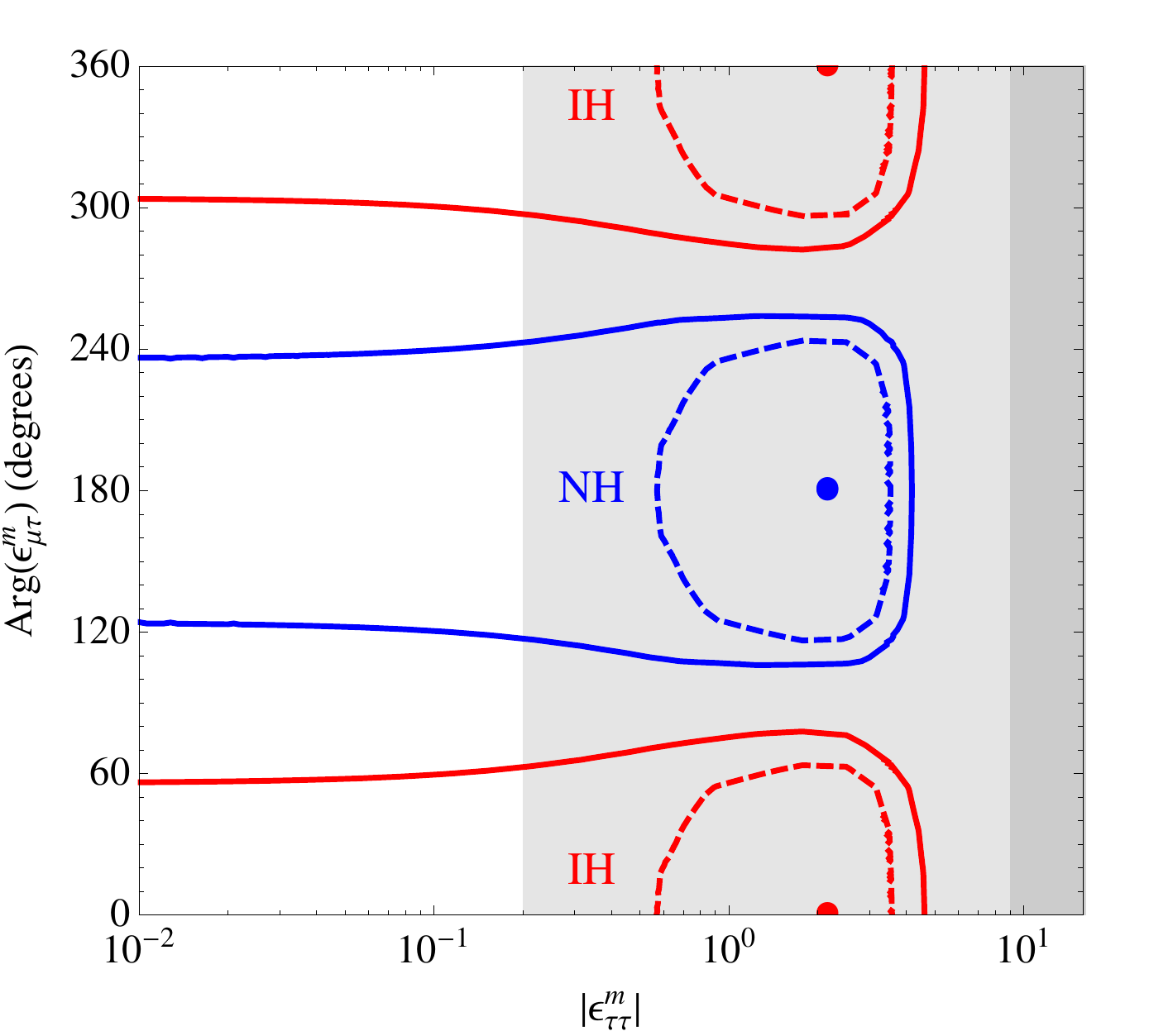}
    \end{tabular}
  \end{center}
  \vspace{-.5cm}
  \caption{Constraints on the parameter space of neutral current non-standard
    interactions in the $\mu$--$\tau$ sector from MINOS $\nu_\mu$ and
    $\bar{\nu}_\mu$ data. Each panel corresponds to a projection of the
    three-dimensional NSI parameter space along one of its axes. This means
    that in each panel we have marginalized over one of the NSI parameters (as
    well as the six standard oscillation parameters).  The best fit points are
    indicated by the colored dots. The symmetries from eqs.~\eqref{eq:symm1-NC}
    and \eqref{eq:symm2-NC} are clearly visible in the plot, while the
    additional two-fold ambiguity eq.~\eqref{eq:symm3-NC} is implicit. In the
    bottom panels, we explicitly indicate the parameter regions corresponding
    to a normal mass hierarchy (NH) and to an inverted mass hierarchy (IH),
    while in the top panel, NH and IH contours lie on top of each other.
    Exclusion limits from other experiments~\cite{Biggio:2009nt,
    GonzalezGarcia:2007ib} are shown in gray.  See text for caveats pertaining
    to bounds from atmospheric neutrino measurements.}
  \label{fig:nsi-fit-NC}
\end{figure}

The allowed regions and best fit values for $|\epsilon^m_{\mu\tau}|$,
$\arg(\epsilon^m_{\mu\tau})$, and $|\epsilon^m_{\tau\tau}|$ are shown in the
three panels of fig.~\ref{fig:nsi-fit-NC}.  We find that, in spite of the large
best fit value for $|\epsilon^m_{\tau\tau}|$, MINOS is compatible with
$\epsilon^m_{\tau\tau} = 0$ at the 90\% confidence level. The case of pure
standard oscillations, on the other hand, is ruled out at 90\% confidence
level. Note that only a fourfold rather than eightfold degeneracy is visible in
fig.~\ref{fig:nsi-fit-NC} because the transformation \eqref{eq:symm3-NC} only
changes the sign of $\epsilon^m_{\tau\tau}$ and the octant of $\theta_{23}$,
which are not displayed here.

The compatibility of the parameter region favored by MINOS with existing
constraints depends strongly on whether atmospheric neutrino bounds are taken
into account. Atmospheric neutrinos are very sensitive to non-standard matter
effects since they can travel over very long distances inside the Earth before
reaching the detector.  Therefore, the MINOS-favored region of parameter space
is in conflict with the limits derived in refs.~\cite{Fornengo:2001pm,
GonzalezGarcia:2007ib} from MACRO and Super-Kamiokande data.  While these
bounds are numerically very strong, they have been derived in a two-flavor
framework.  It has been called into question whether they would still hold when
standard and non-standard three-flavor effects are taken into
account~\cite{Blennow:2008ym, Friedland:2005vy}.  While we consider it highly
unlikely that three-flavor effects would weaken the atmospheric neutrino bounds
on NC NSI by the amount required to restore compatibility with the
MINOS-favored parameter region, we cannot definitely rule out NC NSI as an
explanation for the MINOS data at this time.

%------------------------------------------------------------------------------
\subsection{Charged current NSI}
\label{sec:fit-CC}
%------------------------------------------------------------------------------

Let us now turn to the investigation of the charged current NSI introduced in
sec.~\ref{sec:analytic-CC}. We fit the MINOS data using a three-flavor analysis
including the NSI parameters $|\epsilon^d_{\tau\mu}|$ and
$\arg(\epsilon^d_{\tau\mu})$, but, as for the neutral current case, we find
that three-flavor effects are small. In particular, the symmetries of the
$\nu_\mu$ survival probability found in sec.~\ref{sec:analytic-CC} for the
two-flavor case are present also in the three-flavor framework. In particular,
this means that there is a continuous family of best fit points (see
eqs.~\eqref{eq:symm1-CC} and \eqref{eq:symm2-CC}, and the continuous
generalization of \eqref{eq:symm2-CC} at the end of
sec.~\ref{sec:analytic-CC}). For reference, we here give one representative
point from this family that is most consistent with existing bounds on
$|\epsilon^d_{\tau\mu}|$:
\begin{align}
  \begin{aligned}
    \epsilon^d_{\tau\mu}  &= 0.12 \, i = 0.12 \, e^{0.5 i \pi} \,,&\qquad
    \sin^2 \theta_{23}    &= 0.41                   \,,&\qquad
    \Delta m_{32}^2       &= +2.74 \times 10^{-3}\ \text{eV}^2 \,,
  \end{aligned}
  \label{eq:bf-CC}
\end{align}
with $\chi^2/\text{dof} = 15.9/15$.  The predicted MINOS event spectra at this
point are shown as the solid blue histograms in fig.~\ref{fig:spectrum}.

\begin{figure}
  \begin{center}
    \includegraphics[width=0.45\textwidth]{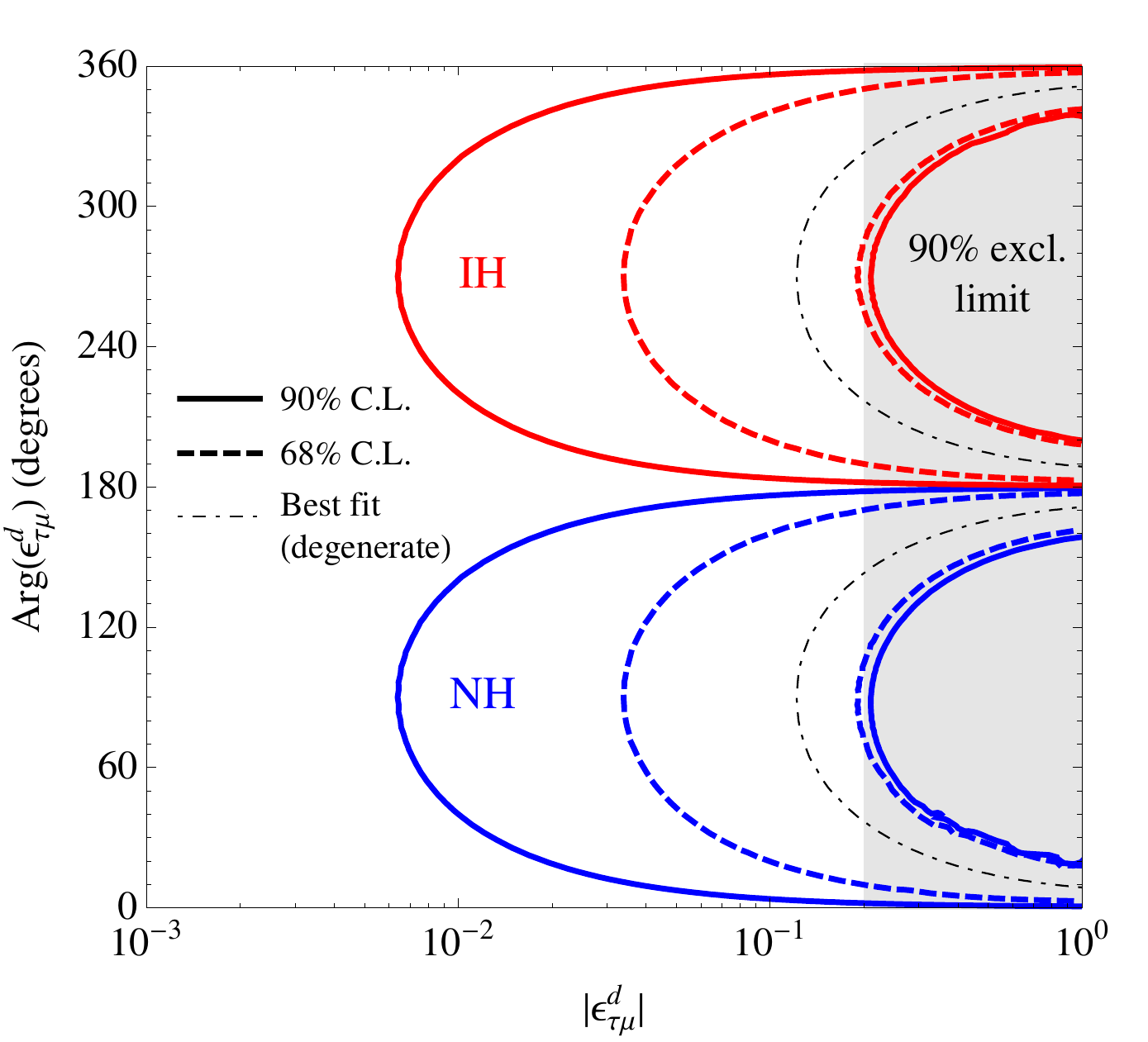}
  \end{center}
  \vspace{-.5cm}
  \caption{Constraints on the parameter space of charged current non-standard
    interactions between $\nu_\tau$ and muons from MINOS $\nu_\mu$ and
    $\bar{\nu}_\mu$ data. We have used a full three-flavor fit, marginalizing
    over the standard oscillation parameters. The discrete and continuous symmetries from
    eqs.~\eqref{eq:symm1-CC} and \eqref{eq:symm2-CC} are clearly visible in the
    plot. The thin dash-dotted black curves indicate the positions of
    the approximately degenerate best fit points.  We explicitly indicate the
    parameter regions corresponding to a normal mass hierarchy (NH, blue) and
    to an inverted mass hierarchy (IH, red). See text for comments on
    existing constraints on CC NSI.}
  \label{fig:nsi-fit-CC}
\end{figure}

In fig.~\ref{fig:nsi-fit-CC}, we show the allowed regions in the
$|\epsilon^d_{\tau\mu}|$--$\arg(\epsilon^d_{\tau\mu})$ plane determined by our
fit. The continuous family of best fit points is indicated by the thin
dash-dotted black curves.  To the best of our knowledge, no bound on the
vector-type $\nu_\tau$--$\mu$ CC NSI considered here exists in the literature.
(The bounds of order 0.1 derived in ref.~\cite{Biggio:2009nt} apply only to
axial-vector operators.) However, the experimental limit on the branching ratio
of the flavor-violating decay $\tau^\pm \to \mu^\pm \pi^0$~\cite{Amsler:2008zz}
can be translated into a bound of order 0.2 on $|\epsilon^d_{\tau\mu}|$ (and,
in fact, also on the related coefficient $|\epsilon^d_{\mu\tau}|$) for vector
and axial-vector type interactions (see appendix~\ref{sec:loopbound} for
details and caveats). We expect that a bound could also be derived from lepton
universality considerations in weak decays of parity-even hadrons, but that it
would not be stronger than $\mathcal{O}(0.1)$. Also, atmospheric neutrinos are
sensitive to $|\epsilon^d_{\tau\mu}|$, but since CC NSI are not enhanced by the
long baselines of atmospheric neutrinos, we expect these constraints to be
relatively weak as well.

%==============================================================================
\section{Testing the NSI interpretation of MINOS data in future experiments}
\label{sec:future}
%==============================================================================

To corroborate or refute the hypothesis of large non-standard interactions as
an explanation for the apparent discrepancy between neutrino and anti-neutrino
results in MINOS, it will be mandatory to gather more statistics in MINOS
itself, and to look for possible NSI signals in future experiments like T2K and
NO$\nu$A. In the following, we will neglect neutral-current NSI since we have
seen in sec.~\ref{sec:fit-NC} that they are disfavored as an explanation for
the MINOS data by atmospheric neutrinos.  Instead, we will focus on CC NSI.  We
have computed the expected event spectrum in MINOS, T2K, and NO$\nu$A assuming
the values of the standard and non-standard oscillation parameters to be given
by the MINOS best fit point eq.~\eqref{eq:bf-CC}. We have then attempted a
standard oscillation (no NSI) fit to this simulated data. If this fit is
incompatible with the simulated data at a given confidence level, we say that
the existence of a non-standard effect can be established experimentally at
this confidence level.  Our simulation of T2K follows~\cite{Huber:2002mx,
Itow:2001ee, Ishitsuka:2005qi}, while that of NO$\nu$A is based
on~\cite{Ayres:2004js, Yang_2004}. We include only the $\nu_\mu$ and
$\bar{\nu}_\mu$ disappearance channels.

\begin{figure}
  \begin{center}
    \includegraphics[width=\textwidth]{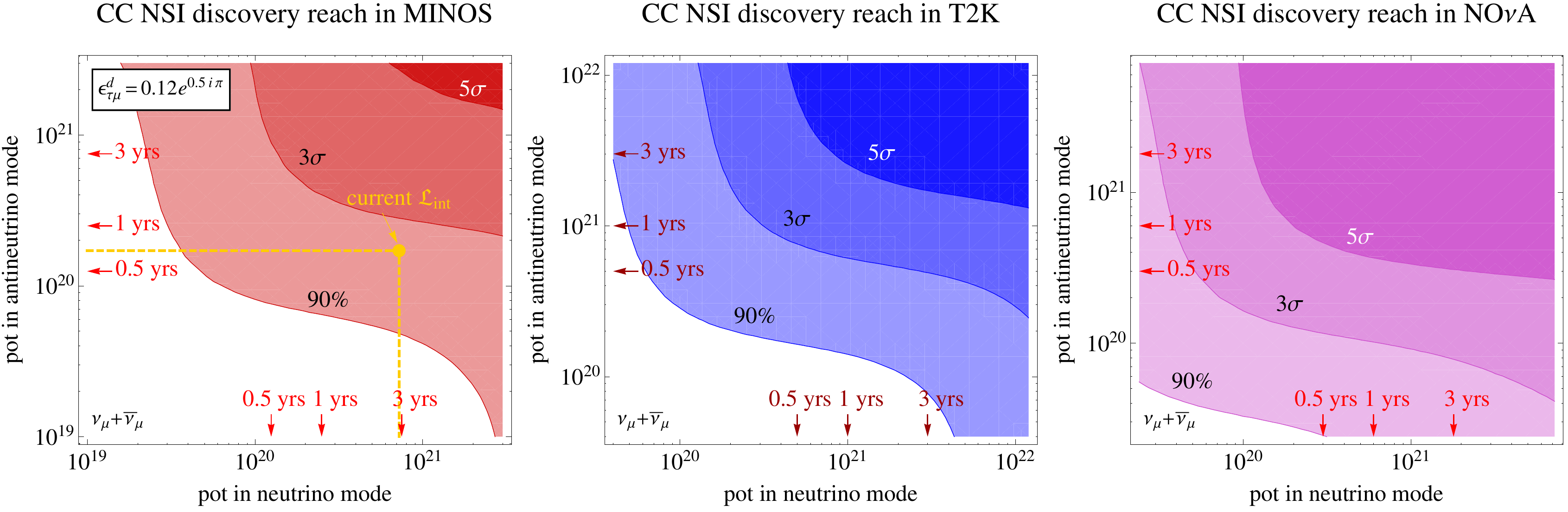}
  \end{center}
  \vspace{-.5cm}
  \caption{Discovery reach for charged current NSI parameters corresponding to
    the MINOS best fit point eq.~\eqref{eq:bf-CC} in MINOS (left), T2K (middle),
    and NO$\nu$A (right).}
  \label{fig:exp2d}
\end{figure}

In fig.~\ref{fig:exp2d}, we show the predicted discovery potential in MINOS,
T2K, and NO$\nu$A as a function of the integrated luminosity in neutrino mode
and the integrated luminosity in anti-neutrino mode. We also indicate how the
number of protons on target (pot) translates into a time of running at nominal
luminosity ($2.5 \times 10^{20}$~pot/year for MINOS, $6 \times
10^{20}$~pot/year for NO$\nu$A, and $10^{21}$~pot/year for T2K). We see that
optimal sensitivity is achieved if slightly more time is spent on running in
anti-neutrino mode than on running in neutrino mode. This is easily
understandable since the assumed NSI effect manifests itself mainly as an
apparent discrepancy between neutrino and anti-neutrino results, while each
data sample individually appears to be consistent with standard oscillations.
Thus, to optimally probe the non-standard effect, the event numbers in the
$\nu_\mu$ and $\bar{\nu}_\mu$ samples should not be too different.  On the
other hand, anti-neutrino cross sections are about a factor of 3 smaller than
neutrino cross section, so more time has to be devoted to $\bar{\nu}$ running
to achieve this goal. Fig.~\ref{fig:exp2d} also shows that in order to improve
the statistical significance of the anomalous effect in MINOS itself, more
anti-neutrino running is desirable since the experiment has already taken a lot
of data in neutrino mode. By comparing the three panels of
fig.~\ref{fig:exp2d}, we see that, as expected, the discovery potential of T2K
is better than that of MINOS, while the best sensitivity is achieved in
NO$\nu$A.  After one year of nominal running in neutrino mode, NO$\nu$A could
confirm the existence of the non-standard effect at the 90\% confidence level,
while in anti-neutrino mode, even a few months would be sufficient to achieve
that sensitivity. This can be understood by noting that, for the parameter
values favored by MINOS, eq.~\eqref{eq:bf-CC}, the two
$\mathcal{O}(|\epsilon^d_{\tau\mu}|)$ NSI terms in eq.~\eqref{eq:P-CC} have
opposite signs for neutrinos, but the same sign for anti-neutrinos. Therefore,
the non-standard effect is stronger for anti-neutrinos.  To achieve a $3\sigma$
discovery in T2K or NO$\nu$A, neutrino \emph{and} anti-running are required,
with at least one year spent in each mode for T2K, or half a year for NO$\nu$A.

\begin{figure}
  \begin{center}
    \includegraphics[width=0.6\textwidth]{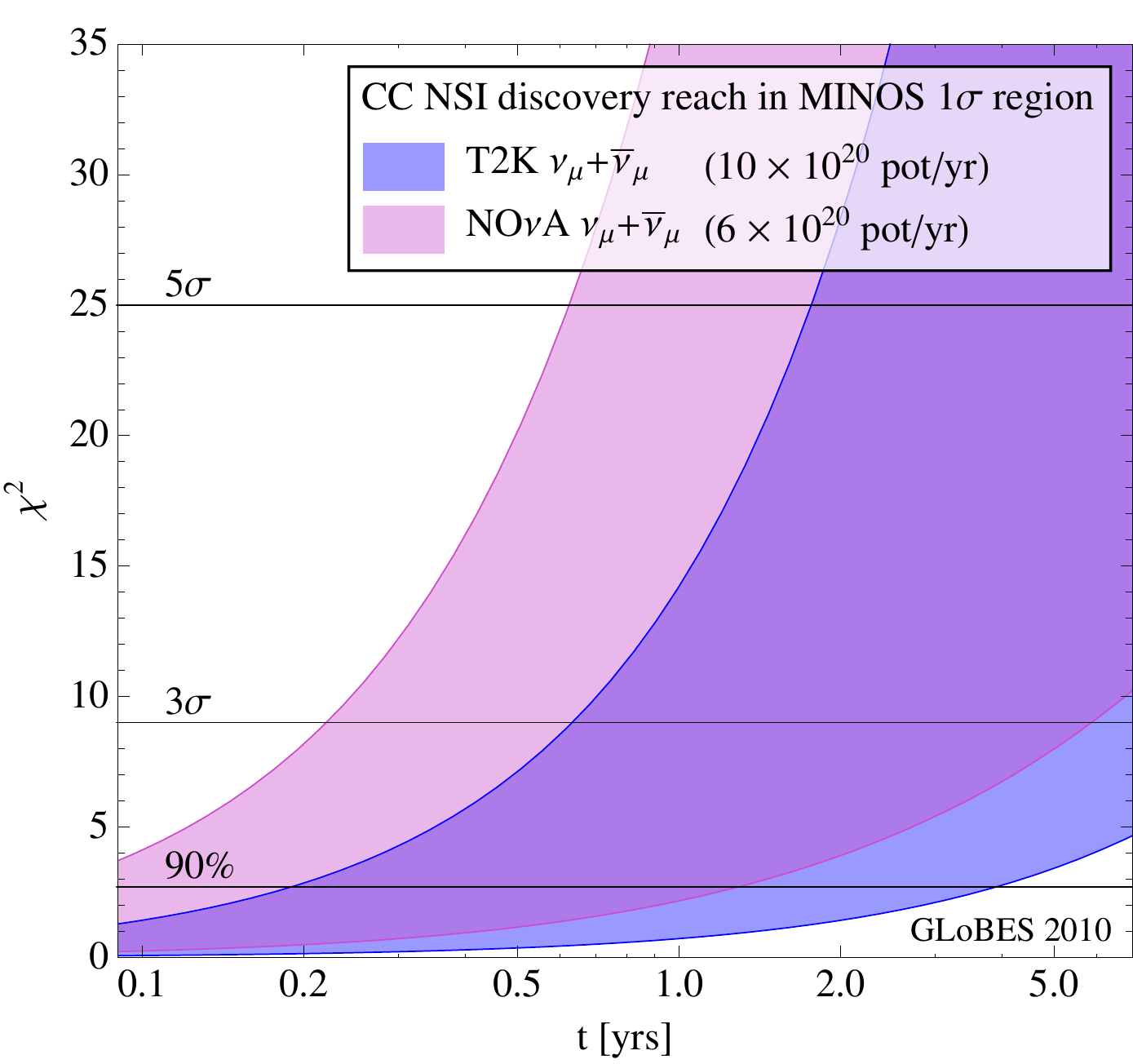}
  \end{center}
  \vspace{-.5cm}
  \caption{$\chi^2$ as a function of exposure time for a standard oscillation
    fit to (simulated) data affected by charged current NSI. The widths of the
    colored bands corresponds to varying the NSI parameters within the $1\sigma$
    allowed region preferred by MINOS (fig.~\ref{fig:nsi-fit-CC}).  We have
    assumed running at the indicated nominal luminosities, and we have assumed
    the running time to be equally divided into neutrino and anti-neutrino
    running.}
  \label{fig:exp-chi2}
\end{figure}

If the true values of the NSI parameters are different from the best fit point
eq.~\eqref{eq:bf-CC}, the discovery reach in future experiments can be altered
significantly. This is illustrated in fig.~\ref{fig:exp-chi2}, where we plot
the $\chi^2$ of a standard oscillation fit to simulated data affected by CC NSI
as a function of the running time at nominal luminosity.  The widths of the
colored bands correspond to the $1\sigma$ uncertainty in the NSI parameters
from fig.~\ref{fig:nsi-fit-CC}.  Fig.~\ref{fig:exp-chi2} shows that if nature has
chosen unfavorable NSI parameters, it will be very hard for T2K and NO$\nu$A to
announce a discovery. On the other hand, for favorable parameter values a
$3\sigma$ effect could be detected after less than one year of nominal running
even in T2K.

%==============================================================================
\section{Discussion and conclusions}
\label{sec:conclusions}
%==============================================================================

We have seen that, in order to explain the tension between the $\nu_\mu$ and
$\bar{\nu}_\mu$ event samples in MINOS using NSI, the NSI couplings would have to be
rather large, almost of the same order as Standard Model weak interactions.
While we have shown in sec.~\ref{sec:fit} that there are regions of parameter
space still consistent with MINOS data and with constraints from other
experiments, one should keep in mind that the effective operators generating
the NSI should ultimately arise from an underlying renormalizable model.
Model-dependent constraints, however, are usually much stronger than the
model-independent bounds we have considered.

For example, the most straightforward implementations of dimension~6 NSI
operators, based on the introduction of new heavy tree-level mediator fields,
are phenomenologically not viable because $SU(2)$ invariance would dictate that
large neutrino NSI realized that way would have to be accompanied by large
non-standard effects in the charged lepton sector~\cite{Antusch:2008tz,
Gavela:2008ra}. Therefore, such models are usually tightly constrained by rare
decay searches~\cite{Amsler:2008zz}. NSI might arise from dimension~8 operators
involving two Standard Model Higgs fields contracted with lepton doublets, so
that after electroweak symmetry breaking, $SU(2)$ breaking 4-fermion couplings
can arise. However, dimension~8 operators of this type are typically
accompanied by phenomenologically problematic dimension~6 operators unless the
coefficients of different operators obey certain cancellation
conditions~\cite{Gavela:2008ra}. Further model-dependent constraints on
neutrino NSI operators can come from electroweak precision tests such as muon
$g-2$ measurements, and from direct searches for possible mediators. All these
constraints will typically force the mediators to be heavy (at least a few
hundred GeV) or very weakly coupled.

The latter possibility---neutrino NSI mediated by light ($\ll M_W$), weakly coupled
particles---is less well explored in the literature, so a scenario of this type
could be responsible for the effects seen by MINOS.  This is particularly
interesting as models containing light new particles have recently received a
lot of attention in the context of Dark Matter searches (see e.g.\ refs.\
\cite{Fitzpatrick:2010em, Hooper:2010uy, Arina:2010rb, Lavalle:2010yw,
Cumberbatch:2010hh, Kuflik:2010ah}).

In conclusion, we have shown that the tension between the $\nu_\mu$ and
$\bar{\nu}_\mu$ disappearance data in MINOS---if it persists---could be
explained by non-standard neutrino interactions. While neutral current
interactions (non-standard matter effects) of the required magnitude are most
likely ruled out by atmospheric neutrino constraints, a charged current
operator leading to flavor-violating couplings between $\tau$-neutrinos and
muons is not excluded.  We have shown that such NSI can be tested in T2K and
NO$\nu$A, provided that the experiments are operated in neutrino \emph{and}
anti-neutrino mode.  It remains an open question if and how NSI large enough to
explain the MINOS results can arise from a renormalizable model.  Along the
way, we have derived the new constraints $|\epsilon^d_{\tau\mu}| < 0.20$ and
$|\epsilon^d_{\mu\tau}| < 0.20$ on flavor-violating vector or axial-vector
type charged current non-standard interactions in the $\mu$--$\tau$ sector by
considering their loop-level contributions to the flavor-violating decay
$\tau^\pm \to \mu^\pm \pi^0$.

\section*{Acknowledgments}

We are indebted to Mary Bishai for providing the simulated NuMI $\nu_\mu$
fluxes used as input to our simulation.  We are also grateful to Bogdan
Dobrescu, Thomas Schwetz-Mangold, and Jure Zupan for several insightful
discussions on model-building aspects of non-standard neutrino interactions,
and to Carla Biggio, Mattias Blennow, Enrique Fernandez-Martinez, Belen Gavela,
and Bill Marciano for very helpful comments and discussions on NSI constraints
from existing data.  Finally, it is a pleasure to thank the organizers of the
Neutrino 2010 conference, at which we learned about the new MINOS results, and
the organizers of the INT Program on Long-Baseline Neutrino Physics and
Astrophysics in Seattle, Washington, during which part of this work was
completed, for two very enjoyable and fruitful meetings. PANM would like to
thank the Funda\c{c}\~{a}o de Amparo \`{a} Pesquisa do Estado de S\~{a}o Paulo
for financial support of his PhD project, and Fermilab for kind hospitality and
support during his visit.  PANM and JK have been partially supported by the US
Department of Energy's Institute for Nuclear Theory (INT) at the University of
Washington. Fermilab is operated by Fermi Research Alliance, LLC under Contract
No.~DE-AC02-07CH11359 with the US Department of Energy.

%==============================================================================
\appendix
\section{GLoBES simulation of MINOS}
\label{sec:sim}
%==============================================================================

\begin{figure}
  \begin{center}
    \includegraphics[width=0.6\textwidth]{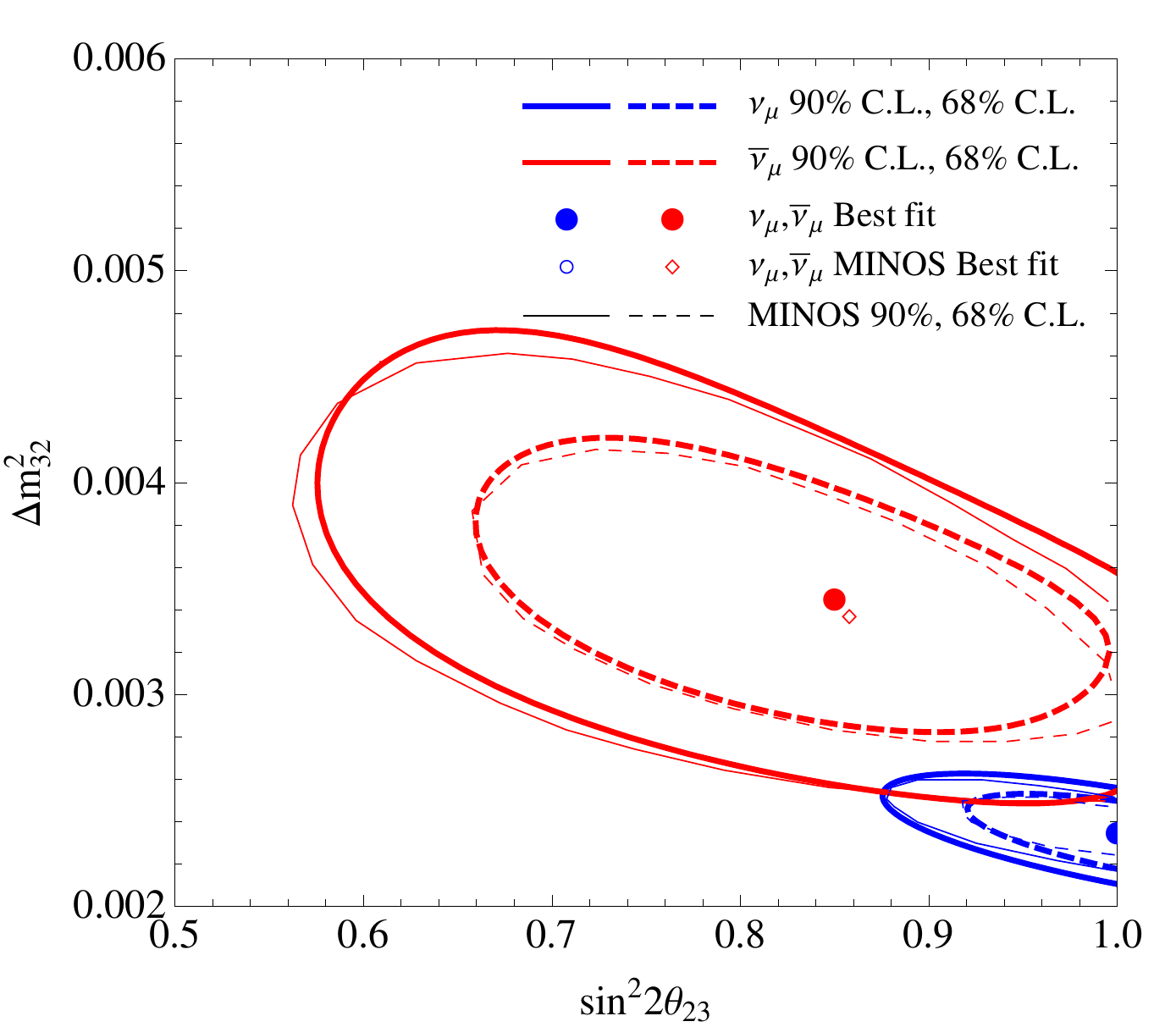}
  \end{center}
  \vspace{-.5cm}
  \caption{Allowed regions in the $\sin^2 2\theta_{23}$--$\Delta m_{32}^2$
    plane from separate two-flavor standard oscillation fits to MINOS
    $\nu_\mu$ and $\bar{\nu}_\mu$ data. The thick contours and filled
    circles represent the 68\% and 90\% contours and best fit points from our
    fit, while thin lines and empty diamonds show the fit
    results obtained by the MINOS collaboration~\cite{MINOS:Nu2010}.}
  \label{fig:std-osc-fit}
\end{figure}

In this appendix, we provide details on the parameters of our MINOS simulation.
We have used GLoBES~\cite{Huber:2004ka, Huber:2007ji}, with an implementation
of NSI developed in refs.~\cite{Kopp:2007mi,Kopp:2007ne}, and with a MINOS
experiment description based on~\cite{Ables:1995wq, Huber:2004ug, NUMIL714,
MINOS:Nu2010}.  The neutrino fluxes are taken from Monte Carlo simulations of
the NuMI beam~\cite{Bishai:privcomm}.  We use the same binning as the MINOS
collaboration, but restrict our analysis to an energy window from 1--5~GeV for
neutrinos, and from 1--8~GeV for anti-neutrinos. The reason for not including
higher-energy neutrino data is that the $\nu_\mu$ fluxes available to us did
not include the effect of unfocused high energy pions in the secondary beam.
Detection efficiencies for $\nu_\mu$, neutral current backgrounds, and the
actual MINOS data were taken from~\cite{MINOS:Nu2010}. The neutrino--nucleon
scattering cross sections are based on the cross sections for water targets
in~\cite{Messier:1999kj,Paschos:2001np}, and to account for the difference
between the cross sections for water and those for iron, we adjust the overall
normalization factor in our simulation in such a way that we optimally
reproduce the predictions of the MINOS Monte Carlo simulation.  We assume a
Gaussian energy smearing function with a width $\sigma_E$ given by $0.16 E +
0.07 \sqrt{E / \text{GeV}}$~GeV for neutrino running, and by $0.155 E + 0.11
\sqrt{E / \text{GeV}}$~GeV for anti-neutrino running.  These numbers were again
optimized to reproduce the event rates predicted by the MINOS Monte Carlo
simulation.  The matter density along the neutrino trajectory is assumed to be
2.8~g/cm$^3$. We have checked that the results of two-flavor standard
oscillation fits to $\nu_\mu$ and $\bar{\nu}_\mu$ data agree well with the
allowed regions obtained by the MINOS collaboration (see
fig.~\ref{fig:std-osc-fit}).

In our fits, we use Gaussian prior terms to impose the $1 \sigma$ constraints
$\sin^2 2\theta_{13} < 0.1$, $\sin^2 \theta_{12} = 0.319 \pm 0.023$, and
$\Delta m_{21}^2 = (7.59 \pm 0.30) \times 10^{-5}$~eV$^2$~\cite{Maltoni:2008ka,
GonzalezGarcia:2010er}.  As systematic errors, we include independent 4\% (3\%)
normalization uncertainties on the signal (background) rates for neutrinos, and
5\% (5\%) normalization uncertainties on the signal (background) rates for
anti-neutrinos.

%==============================================================================
\section{New loop bounds on non-standard interactions}
\label{sec:loopbound}
%==============================================================================

\begin{figure}
  \begin{center}
    \begin{tabular}{c@{\hspace{0.5cm}}c@{\hspace{0.5cm}}c}
      \includegraphics{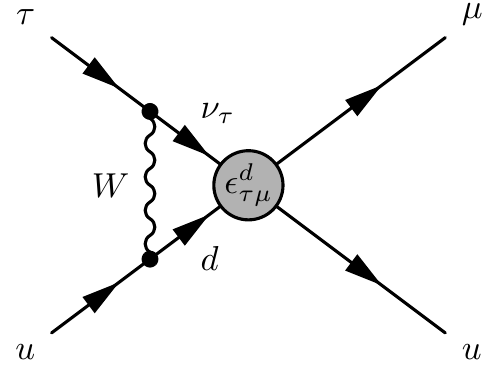} &
      \includegraphics{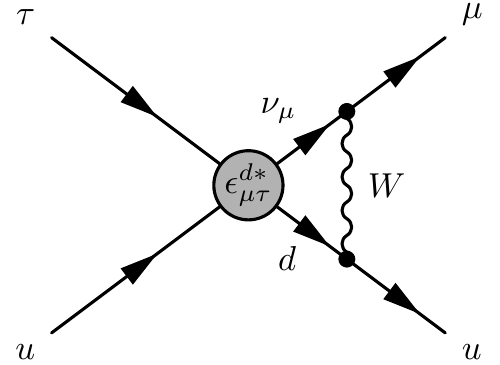} &
      \includegraphics{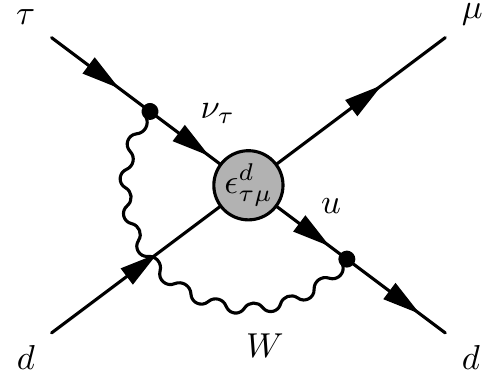} \\[0.4cm]
      (a) & (b) & (c)
    \end{tabular}
  \end{center}
  \caption{(a) and (b) The loop diagrams used to constrain $\epsilon^d_{\tau\mu}$
    and $\epsilon^d_{\mu\tau}$, respectively. (c) A similar diagram that does
    not have a logarithmic divergence.}
  \label{fig:loop1}
\end{figure}

Here, we outline the calculation of the loop diagram fig.~\ref{fig:loop1}~(a) which
we use to translate the experimental bound on the rare decay $\tau^\pm \to \mu^\pm
\pi^0$ into a constraint on $|\epsilon^d_{\tau\mu}|$. We perform the
calculation in unitary gauge and keep only the logarithmically divergent part.
Indeed, it has been argued in refs.~\cite{Davidson:2003ha, Biggio:2009kv,
Biggio:2009nt} that the finite and quadratically divergent terms cannot be
calculated in a model-independent way, while the logarithmic divergence
corresponds to the model-independent one-loop renormalization group running of
the effective NSI operator from the UV completion scale down to $M_W$.
(This argument is invalid if the UV complete theory leads to another
effective operator having the same renormalization group running and exactly
canceling the NSI operator. If that case is considered, no model-independent
bound on $|\epsilon^d_{\tau\mu}|$ can be derived.)
Neglecting fermion masses, we find that the logarithmically divergent part of
fig.~\ref{fig:loop1}~(a) is
\begin{align}
  \frac{3\sqrt{2} G_F \epsilon^d_{\tau\mu} V_{ud} \, \alpha}{2\pi s_w^2}
  \log \frac{\Lambda}{M_W} \big[ \bar{\tau} \gamma^\mu P_L \mu \big]
  \big[ \bar{u} \gamma^\mu P_L u \big] \,,
  \label{eq:loop1}
\end{align}
where $\alpha$ is the electromagnetic fine structure constant, $s_w$ is the sine
of the weak mixing angle, $V_{ud}$ is a CKM matrix element, $P_L = (1 - \gamma^5)/2$,
and $\Lambda$ is the UV
completion scale. Comparing eq.~\eqref{eq:loop1} to the operator
$2 \sqrt{2} G_F V_{ud} \, [ \bar{\tau} \gamma^\mu P_L \nu_\tau]
[ \bar{u} \gamma^\mu P_L d]$ responsible for the standard decay
$\tau^\pm \to \pi^\pm \nu_\tau$, using the QCD isospin symmetry between $\pi^0$
and $\pi^\pm$, and assuming $\log \Lambda/M_W \sim \mathcal{O}(1)$,
we find
\begin{align}
  |\epsilon^d_{\tau\mu}| \simeq
    \sqrt{\frac{2 \, \text{BR}(\tau^\pm \to \mu^\pm \pi^0)}
         {\text{BR}(\tau^\pm \to \pi^\pm \nu_\tau)}} \,
         \frac{4 \pi s_w^2 }{3 \alpha} \,.
  \label{eq:epsbound-1}
\end{align}
The factor 2 below the square root is due to the fact that the $\pi^0$ can be
interpreted as a state $(\bar{u} u - \bar{d} d)/\sqrt{2}$ whose overlap with
the final state $\bar{u} u$ from fig.~\ref{fig:loop1}~(a) is only $1/\sqrt{2}$.
With the experimental constraint $\text{BR}(\tau^\pm \to \mu^\pm \pi^0) < 1.1
\times 10^{-7}$~\cite{Amsler:2008zz}, eq.~\eqref{eq:epsbound-1} leads to the
new bound
\begin{align}
  |\epsilon^d_{\tau\mu}| \lesssim 0.20 \,.
  \label{eq:epsbound-2}
\end{align}
In full analogy to the above, we can also use fig.~\ref{fig:loop1}~(b)
to set a bound on $\epsilon^d_{\mu\tau}$:
\begin{align}
  |\epsilon^d_{\mu\tau}| \lesssim 0.20 \,.
  \label{eq:epsbound-3}
\end{align}
Note that diagrams like the one in fig.~\ref{fig:loop1}~(c) with $\bar{d} d$ in
the final state do not contribute to these bound because they do not
have logarithmic divergences.

While we have derived eqs.~\eqref{eq:epsbound-2} and \eqref{eq:epsbound-3} by
assuming vector-type NSI, our calculations can be easily generalized to set
equivalent bounds on axial-vector operators and on $V-A$ operators, while no
constraints can be derived for $V+A$ type interactions since the diagrams in
fig.~\ref{fig:loop1} are suppressed by a neutrino mass insertion in this case.

% Bibliography

\end{document}